\documentclass[oldversion]{aa}
\usepackage[numbers]{natbib}
\usepackage{graphicx}
\def\etal{{\it et al.\ }}

\def\kms{\ {\rm km\,s^{-1}}}

\begin{document}

\title{New measurements of radial velocities in clusters of galaxies-V
\thanks{based on observations made Haute-Provence observatory (France)}}
\author{D.~Proust\inst{1} \and H.V. Capelato\inst{2} 
\and G.B.~Lima~Neto\inst{3} \and L.~Sodr\'e Jr.\inst{3}}

\offprints{D.~Proust} 

\institute{%
Observatoire de Paris-Meudon, GEPI, F92195 MEUDON, France 
\and Divis\~ao de Astrof\'{\i}sica, INPE/MCT, 12227-010, 
     S\~ao Jos\'e dos Campos/S.P., Brazil
\and Instituto de Astronomia, Geof\'{\i}sica e Ci\^encias Atmosfericos, 
     Universidade de S\~ao Paulo (IAG/USP), 05508-090 S\~ao Paulo/S.P.,
Brazil}

\date{Received date; accepted date}

\abstract{As a part of our galaxy-cluster redshift survey, we
present a set of 80~new velocities in  the 4~clusters Abell~376,
Abell~970, Abell~1356, and Abell~2244, obtained at Haute-Provence
observatory. This set now completes our previous analyses, especially for 
the first two clusters. Data on individual galaxies are presented, and we
discuss some cluster properties. For A376, we obtained an improved mean
redshift $\overline{z} = 0.047503$ with a velocity dispersion of
$\sigma_V = 860\kms$. For A970, we have $\overline{z} = 0.058747$ with
$\sigma_V = 881\kms$. We show that the A1356 cluster is not a member
of the ``Leo-Virgo'' supercluster at a mean redshift $\overline{z}=
0.112$ and should be considered just as a foreground group of galaxies
at $\overline{z} = 0.0689$, as well as A1435 at $\overline{z} = 0.062$.
We obtain $\overline{z} = 0.099623$ for A2244 with  $\sigma_V = 965\kms$.
The relative proximity of clusters A2244 and A2245 ($\overline{z} =
0.0873816$, $\sigma_V = 992\kms$) suggests that these could be members of
a supercluster that would include A2249; however, from X-ray
data there is no indication of interaction between A2244 and A2245.

}

\maketitle

\section{Introduction} \label{Introduction}

Redshift surveys in clusters of galaxies are needed to study their
dynamical and evolutionary state. In clusters, the mean velocity is a 
key factor in deriving distances, allowing the study of matter
distribution on very large scales. Within clusters analysis of 
the velocity field can lead to an estimate of the virial mass,
constraining models of the dark matter content. Galaxy velocity
measurements provide informations that are complementary to other 
wavelengths, in particular what are obtained through X-ray observations
of clusters. Optical, spectroscopic, and X-ray data form basic pieces of
information for the mass estimates. However, discrepancies between these
estimators are often found (e.g. Girardi \etal 1998, Allen 2000, Cypriano
\etal 2005). Virial mass estimates rely on the assumption of dynamical
equilibrium. X-ray mass estimates also depend on the dynamical
equilibrium hypothesis and on the still not well-constrained
intracluster gas temperature gradient (e.g., Leccardi \& Molendi 2008). 
Finally, mass estimates based on gravitational lensing are
considered more reliable than the others (e.g., Mellier 1999) because
they are completely independent of the dynamical status of the cluster.
The drawback is that lensing can only probe the central region of
clusters. The discrepancies among the methods may come from the
non-equilibrium effects in the central region of the clusters (Allen,
1998).

In this paper, we complete our preceeding studies of the dynamical status
of the two clusters Abell~376 (Proust \etal 2003) and Abell~970
(Sodr\'e \etal 2001, Lima Neto \etal 2003) with the addition of 46 and
14~galaxies, respectively. The observations of radial velocities
reported here are part of a program to study the dynamical structure of
clusters of galaxies, which was started years ago and which had several
already published analyses (see e.g. Proust \etal 1992, 1995, 2000;
Capelato \etal 1991, 2008 and references above).

We have added only 10~galaxies in each of the two clusters Abell~1356 and
Abell~2244 since a larger set of velocities in these two clusters have
been obtained in the course of the Sloan Digital Sky Survey
(SDSS)\footnote{http://www.sdss.org/}. For that reason spectroscopic
observations were no longer pursued in these two clusters.

We present in Sect.~2 the details of the observations and data reduction. 
In Sect.~3 we discuss the distribution and the velocity analysis of 
the cluster galaxies, and we summarize our conclusions for each cluster.
We adopt here, whenever necessary, $H_{0}= 70\, h_{70}\kms$Mpc$^{-1}$,
$\Omega_{M} = 0.3$ and $\Omega_{\Lambda} = 0.7.$

\section{Observations and data reductions}\label{Observations and Data
Reductions}

The new velocities presented in this paper were obtained with the
1.93m telescope at Haute-Provence Observatory. Observations were carried
out in April~2000, May~2001 and January~2005. We used the CARELEC
spectrograph at the Cassegrain focus, equipped with a 150~line/mm 
grating blazed at 5000~{\AA} and coupled to an EEV CCD detector
2048x1024 pixels with a pixel size of 13.5~$\mu$m. A dispersion of 
260~{\AA}/mm was used, providing spectral coverage from 3600 to
7300~{\AA}. Wavelength calibration was done using exposures of He-Ne
lamps.

The data reduction was carried out with IRAF\footnote{IRAF is distributed
by the National Optical Astronomy Observatories, which are operated by the
Association of Universities for Research in Astronomy, Inc., under
cooperative agreement with the National Science Foundation.} using the
LONGSLIT package. Radial velocities were determined using the
cross-correlation technique (Tonry and Davis 1979) implemented in the
RVSAO package (Kurtz \etal 1991, Mink \etal 1995) with radial velocity
standards obtained from observations of late-type stars and previously
well studied galaxies.

A total of 80~velocities was obtained from our observations.
Table~1 (available in electronic form at the CDS
via anonymous ftp 130.79.128.5) lists positions and heliocentric velocities for individual galaxies with the following columns:

\begin{enumerate}
\item number of the object. For A376, this number refers to Dressler (1980)
and for A970 it continues the list of Sodr\'e \etal (2001);
\item right ascension (J2000);
\item declination (J2000);
\item morphological type either from Dressler's (1980) catalog for A376
and from a visual inspection on the Palomar Sky Survey (POSS) for A970;
\item heliocentric radial velocity with its error in $\kms$;
\item R-value derived from Tonry \& Davis (1979);
\item notes.
\end{enumerate}

We searched  in the NED  database\footnote{The NASA/IPAC Extragalactic
Database (NED) is operated by the Jet Propulsion Laboratory, California
Institute of Technology, under contract with the National Aeronautics
and Space Administration.} for additional velocities to complement our
redshift samples. As mentioned before, the fields of A1356 and A2244 are
within the sky coverage area of the Sloan Digital Sky Survey, and because
of that most of the 20~measurements made for these clusters resulted in
duplicated data: only 5~new redshifts could have contributed to A1356 and
none for A2244. This last cluster was studied by Rines and Diaferio (2006)
using the caustics technic (Diaferia, 1999) to remove interlopers and to
estimate the velocity dispersions within $r_{200}$. We have no more
information to add to their work.

\makeatletter\if@referee\renewcommand\baselinestretch{1.0}\fi\makeatother
\begin{table*}[htb]
\caption[]{Heliocentric redshift, position, and morphological type for
galaxies of A376, A970, A1356, and A2244.}
\begin{tabular}{lllllll}
\hline
\hline
Galaxy &  R.A.  &  Decl. & Type & Hel. Vel. & R & N \\
 id.   & (2000) & (2000) &    & $V {\pm {\Delta}V}$  & & \\
\hline
{\bf A376} &            &           &        &         &     & \\
  1 & 02 46 29.0 & +36 27 19 &    S   & 40451  32 &  6.53 & 1  \\
  4 & 02 46 21.0 & +36 30 43 &    E   & 39967  26 &  5.93 & 2  \\
 10 & 02 46 40.5 & +36 35 44 &    S0  & 36533  49 &  3.08 &    \\
 12 & 02 46 25.4 & +36 34 48 &    S0  & 13881  59 &  3.98 &    \\
 13 & 02 46 25.3 & +36 32 24 &    S   & 15026  61 &  5.69 & l1 \\
 14 & 02 45 47.0 & +36 35 57 &    S   & 13526 127 &  3.05 &    \\ 
 15 & 02 45 07.6 & +36 36 49 &    S0  & 14018  74 &  5.03 &    \\
 16 & 02 45 06.7 & +36 34 02 &    S   & 12977  68 &  4.49 & 3  \\
 17 & 02 44 43.9 & +36 35 46 &  E/S0  & 13638 101 &  3.22 &    \\
 19 & 02 46 47.6 & +36 40 38 &    S0  & 13629 108 &  3.01 & weak \\
 21 & 02 45 55.6 & +36 43 14 &    S0  & 14873  53 &  3.68 & l2 \\
 23 & 02 45 32.6 & +36 40 46 &    S   & 13398  29 & 10.65 & 4  \\
 29 & 02 44 52.5 & +36 40 21 &    S   & 15602  68 &  3.35 &    \\ 
 30 & 02 44 16.2 & +36 44 00 &    S   & 41071  50 &  3.79 & 5  \\
 38 & 02 46 25.0 & +36 48 37 &    E   & 13975 131 &  3.76 &    \\
 39 & 02 46 20.5 & +36 46 20 &  E/S0  & 15415 100 &  3.39 &    \\
 46 & 02 44 58.9 & +36 45 54 &    S   & 14401  49 &  4.25 & l3 \\
 47 & 02 44 53.9 & +36 47 32 &    Ep  & 14514  60 &  3.29 & l4 \\
 48 & 02 44 44.5 & +36 45 00 &    I   & 14702  56 &  6.69 & 6  \\
 50 & 02 44 11.1 & +36 45 36 &   S0   & 29935  61 &  4.25 & l5 \\
 51 & 02 43 49.1 & +36 46 30 &  S0/E  & 41546  51 &  5.49 &    \\
 53 & 02 47 31.8 & +36 49 45 &    S   & 13701  21 &  8.87 &    \\
 56 & 02 46 54.5 & +36 53 03 &    E   & 13897  58 &  4.69 & l6 \\
 58 & 02 46 34.2 & +36 54 19 &    S   & 16215  66 &  4.00 & l7 \\
 59 & 02 46 28.3 & +36 51 55 &    S   & 13310  94 &  3.49 &    \\
 62 & 02 46 19.7 & +36 50 52 &   S/I  & 13772  73 &       & 7  \\ 
 74 & 02 45 48.7 & +36 52 28 &   S0   & 13466  86 &  3.02 & weak \\
 77 & 02 45 41.1 & +36 52 39 &    S   & 13553  65 &  3.03 &    \\
 79 & 02 45 33.5 & +36 51 43 &   S0   & 14184 119 &  3.19 &    \\
 80 & 02 45 28.8 & +36 53 33 &   S0   & 14948 118 &  3.02 &    \\
 82 & 02 45 22.6 & +36 50 32 &  S0/S  & 26848  58 &  3.73 &    \\
 83 & 02 45 16.5 & +36 50 45 &   S0   & 15265  35 & 12.35 & 8  \\
 87 & 02 44 16.6 & +36 54 49 &   S0   & 12072  78 &  4.03 & 9  \\
 99 & 02 45 47.9 & +36 59 09 &   S0   & 13421  63 &  6.41 &    \\
105 & 02 45 09.4 & +37 00 57 &   S0   & 13550  98 &  3.81 &    \\
107 & 02 43 54.1 & +36 57 49 &  E/S0  & 14365  81 &  4.28 &    \\
110 & 02 45 41.9 & +37 06 05 &    E   &  9255  49 &  5.96 & l8 \\
112 & 02 43 51.6 & +37 03 30 &    S   & 13535  88 &  2.54 & weak \\
114 & 02 46 08.5 & +37 09 49 &    S   & 13260  98 &  3.04 & 10 \\
115 & 02 44 56.2 & +37 11 56 &    S   & 44634  76 &  3.02 &    \\
116 & 02 44 14.2 & +37 09 24 &  E/S0  & 40204  75 &  5.16 &    \\
117 & 02 44 18.8 & +37 08 20 &    S0  & 40124  74 &  3.10 &    \\
118 & 02 45 38.4 & +37 18 18 &    S   & 17561  66 &  4.32 & 11 \\
119 & 02 45 38.7 & +37 18 25 &  S0/a  & 14587  55 &  4.29 & l9 \\ 
120 & 02 45 07.8 & +37 14 24 &    E   & 14530  53 &  4.49 &    \\
\hline
\end{tabular}
\end{table*}
\begin{table*}[htb]
\begin{tabular}{lllllll}
\hline
\hline
Galaxy &  R.A.  &  Decl. & Type & Hel. Vel. & R & N \\
 id.   & (2000) & (2000) &    & $V {\pm {\Delta}V}$  & & \\
\hline
{\bf A970}  &            &           &        &           &  & \\
 70 & 10 16 05.3 & -10 57 40 &    E   & 17014 112 &  3.47 &    \\
 71 & 10 16 08.0 & -10 29 49 &  E/S0  & 11033  91 &  5.70 & 12 \\
 72 & 10 16 10.8 & -10 28 28 &   S0   & 55419  77 &  2.96 & weak \\
 73 & 10 16 14.0 & -10 31 13 &   S0   & 17610 116 &  3.07 & l10 \\
 74 & 10 16 25.7 & -10 57 43 &  E/S0  & 18009 154 &  3.01 & l11 \\
 75 & 10 16 31.9 & -10 37 45 &   S0   & 12360  93 &  3.66 & l12 \\
 76 & 10 16 39.8 & -10 59 59 &    E   & 15662  94 &  6.18 & l13 \\
 77 & 10 16 41.3 & -10 59 03 &  S0/S  & 47793  60 &  3.04 &    \\ 
 78 & 10 17 17.5 & -11 07 50 &   S0   & 17709  83 &  4.26 &    \\
 79 & 10 18 53.9 & -10 49 25 &    S   & 19173  17 &       & 13 \\
 80 & 10 18 59.0 & -10 54 02 &    E   &  2701  23 &       & 14 \\
 81 & 10 19 09.6 & -10 58 39 &    S   & 19404  32 &       & 15 \\
 82 & 10 19 13.7 & -10 22 31 &    E   & 33248  50 &  3.32 &    \\
 83 & 10 19 16.2 & -10 22 09 &  S0/S  & 16887 111 &  3.01 & weak \\
    &            &           &        &           &       &    \\
{\bf A1356} &            &           &        &           &  & \\
  1 & 11 42 04.8 & +10 21 49 &    S   &  6395  43 &  5.19 & l14 \\
  2 & 11 42 08.7 & +10 27 19 &    S   & 21030  82 &  3.63 &    \\     
  3 & 11 42 15.3 & +10 26 50 &    E   & 35166  27 & 11.75 & l15 \\
  4 & 11 42 22.3 & +10 44 19 &    S   & 23753  45 &  8.93 & l16 \\
  5 & 11 42 23.7 & +10 26 09 &    E   & 21474  54 &  6.96 & 16  l29 \\
  6 & 11 42 24.4 & +10 40 04 &    S   & 23794  79 &  6.70 & l17 \\
  7 & 11 42 29.6 & +10 29 24 &    S   & 20910  75 &  6.27 &    \\    
  8 & 11 42 29.7 & +10 28 31 &   S0   & 20702  51 & 10.56 & l18 \\
  9 & 11 42 31.7 & +10 25 24 &    S   & 10655  75 &  5.61 & 17  l30\\
 10 & 11 42 43.4 & +10 31 00 &    E   & 36132  74 &  3.62 & weak  l31 \\
    &            &           &        &        &          &    \\
{\bf A2244} &            &           &        &           &  & \\
  1 & 17 02 16.7 & +33 58 49 &    S   & 29626  39 & 10.37 & l19 \\
  2 & 17 02 25.9 & +33 59 54 &   S0   & 29062  32 & 11.18 & l20 \\
  3 & 17 02 42.5 & +34 03 38 &   S0   & 29811  46 &  9.50 & l21 \\
  4 & 17 02 45.6 & +34 03 39 &    ?   & 27984  66 &  8.67 & l22 \\
  5 & 17 03 23.5 & +34 06 49 &    S   & 11080  25 & 12.40 & 18 l23 \\
  6 & 17 04 03.3 & +33 55 19 &    S   & 11144  39 & 10.96 & l24 \\
  7 & 17 04 04.1 & +34 15 29 &   S0   & 30606  73 &  6.83 & l25 \\
  8 & 17 04 14.4 & +34 15 03 &    S   & 12180  80 &  6.39 & 19 l26 \\
  9 & 17 04 45.9 & +34 06 34 &   S0   & 30707  58 &  8.76 & l27 \\
 10 & 17 04 50.1 & +34 05 29 &    S   &  9959  58 &  7.29 & 20 l28 \\ 

\hline
\end{tabular}
\begin{flushleft}
\smallskip
Data from NED: \\
\textbf{l1} $15130\kms$, \textbf{l2} $14958\kms$, \textbf{l3} $14590\kms$,
\textbf{l4} $14466\kms$, \textbf{l5} $29892\kms$, \textbf{l6} $13988\kms$,
\textbf{l7} $16273\kms$, \textbf{l8} $9324\kms$, \textbf{l9} $14765\kms$,
\textbf{l10} $17631\kms$, \textbf{l11} $18067\kms$, \textbf{l12} $12098\kms$, 
\textbf{l13} $15782\kms$, \textbf{l14} NGC3819 $6274\kms$,
\textbf{l15} $35155\kms$, \textbf{l16} $23768\kms$, \textbf{l17} $23789\kms$,
\textbf{l18} $20773\kms$, \textbf{l19} $29585\kms$, \textbf{l20} $29046\kms$, 
\textbf{l21} $29652\kms$, \textbf{l22} $27955\kms$, \textbf{l23} $11121\kms$, 
\textbf{l24} $11156\kms$, \textbf{l25} $30655\kms$, \textbf{l26} $12110\kms$, 
\textbf{l27} $30809\kms$, \textbf{l28} $9908\kms$, \textbf{l29} $21288\kms$, 
\textbf{l30} $10601\kms$, \textbf{l31} $36047\kms$ \\
\smallskip
Notes: \\
\textbf{1}  emission line:  $H\alpha= 40498\kms$,  \textbf{2} emission
line:  $H\alpha=  39982\kms$,   \textbf{3}  emission  line:  $H\alpha=
13072\kms$,   \textbf{4}  emission  line:   $H\beta$,  2O\textsc{iii},
$H\alpha,  S2$,   \textbf{5}  emission  line:   $H\alpha=  41167\kms$,
\textbf{6}  emission line:  $H\alpha= 14829\kms$,  \textbf{7} emission
line:   O\textsc{ii},   $H\beta$,   2O\textsc{iii},   $H\alpha,   S1$,
\textbf{8}  emission   line:  $H\alpha,  S2=   15478\kms$,  \textbf{9}
emission  line:  $H\alpha=   12066\kms$,  \textbf{10}  emission  line:
$H\alpha= 13267\kms$, \textbf{11} emission line: $H\alpha= 17652\kms$,
\textbf{12} emission line:  $H\alpha= 11058\kms$, \textbf{13} emission
line: $H\alpha$, \textbf{14} emission lines: 2O\textsc{iii}, $H\alpha,
S1$, \textbf{15} emission line: $H\alpha$, \textbf{16} emission lines:
O\textsc{ii},  O\textsc{iii},  $H\alpha$,   $N2=  21398  \pm  68\kms$,
\textbf{17}  emission  lines:   2O\textsc{iii},  $H\alpha=  10684  \pm
35\kms$,  \textbf{18}  emission   lines:  $H\alpha,  N2$,  \textbf{19}
emission  line:  $H\alpha$,   \textbf{20}  emission  lines:  $H\beta$,
2O\textsc{iii}, $H\alpha, N1, N2, S1, S2$ \\

\end{flushleft}
\end{table*}
\makeatletter\if@referee\renewcommand\baselinestretch{1.5}\fi\makeatother

For already observed galaxies, velocity comparison was made between our
data set and NED. We obtained $<V_{o}-V_{\rm ref}> = 31\kms$, the
standard deviation of the difference being $67\kms$. These results are
consistent with the errors of Table~1. The velocities in the present
study agree with those previously published within the $2\sigma$
level.

\section{Galaxy distribution and kinematical analysis} \label{Galaxy disvel}


\subsection{Abell~376}

\begin{figure}[!htb]
\centering
\includegraphics[width=8.6cm]{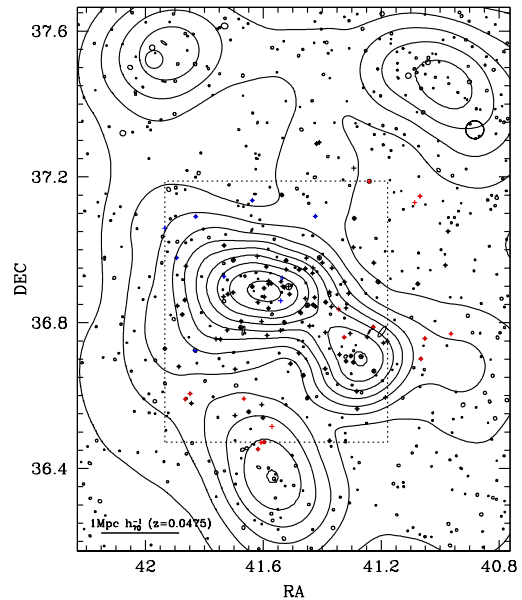}
    \caption{Adaptative kernel density map of the 168 galaxies 
    brighter than $O = 18$ projected in the field of Abell~376. The 
    positions of galaxies brighter than $O = 19.5$ are plotted as 
    ellipses of diameters, ellipticities and position angles taken 
    from the APS catalog. Galaxies having measured redshifts are 
    marked with pluses: blue for foreground galaxies and red for 
    background ones.}
    \label{fig:A376map}
\end{figure}

When including previous measurements (Proust \etal 2003, hereafter P03),
there is a total of 113~measured velocities in the field of Abell~376, from
which 40~are new ones. Note that galaxies from our redshift sample were
selected from the the morphological sample of A376 by Dressler (1980), being
almost complete (113 out of 120). As in P03, galaxy photometry were provided
by the P0SS~I Revised APS Catalogue\footnote{The POSS-I Revised APS Catalogue
is available at the MAPS database from the University of Minnesota, at
http://aps.umn.edu/}(Cabanela et al, 2003) which  gives integrated magnitudes
in the blue photographic $O$ band. Figure \ref{fig:A376map}, which is equivalent
to Figure~1 of P03, shows the projected distribution of galaxies in the field of
Abell~376, where galaxies with measured redshifts have been identified.

We estimated the completeness level of the redshift sample by considering the
minimum rectangular area subtending the entire redshift sample. We find that
the completeness reaches a  maximum of only 57\% at 18~mag. If considering,
however, the same (rectangular) central region previously studied in P03 (see
Figure \ref{fig:A376map}), the completeness level increases to a maximum of
67\% at 18~mag, which is more acceptable. In view of this, we restrain the
kinematical analysis to the same central region as was studied in P03. Moreover,  
as  discussed in P03, despite its incompleteness, the 19.5~mag sample  
was found to be representative of the spatial distribution of galaxies in the central region of the cluster. This allows a more detailed kinematical analysis of central region of the cluster.

\begin{figure}[htb] \centering
\includegraphics[width=8.6cm,angle=-90]{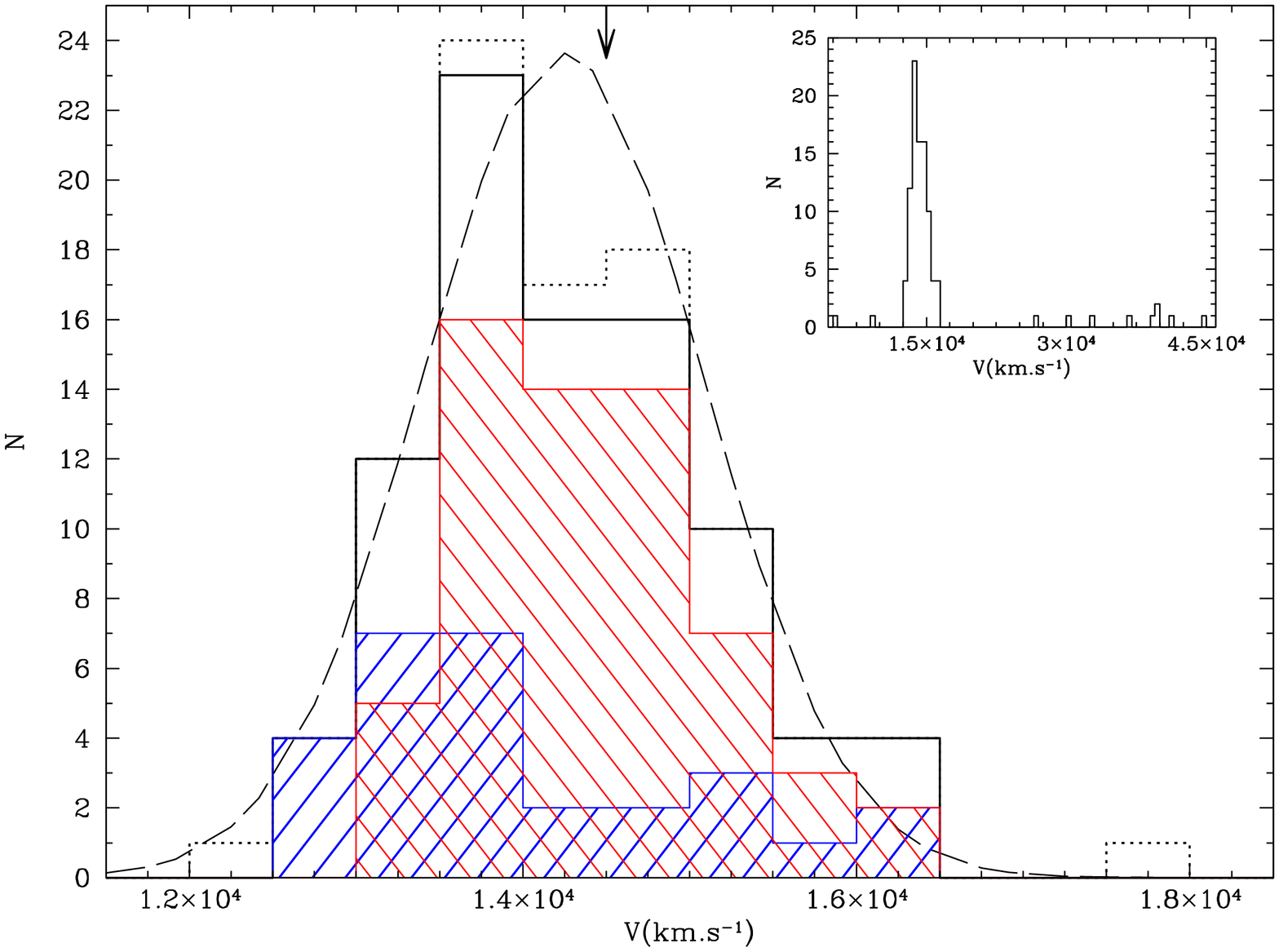}
\caption[]{The radial  velocity distribution for  the Abell~376 sample
of galaxies  between $12000$ and $18000\kms$ contained  in the central
region indicated  in Figure \ref{fig:A376map} . The  dashed line curve
shows the Gaussian distribution corresponding to the mean velocity and
velocity  dispersion quoted  in  the text  (normalized  to the  sample
size).  The inset  show the distribution for the  entire sample of 113
redshifts in  the region  of A376 (dotted  line histogram in  the main
figure).  The  arrow indicates the position  of a gap  in the velocity
distribution  (see  text).  The  shaded  histograms  show  the  radial
velocity distributions  of the  E+S0 galaxies (left-handed  shade; red
lines) and for the S+I galaxies (right hand shade; blue lines).}
\label{fig:A376histov}
\end{figure}

We used the ROSTAT routines (Beers \etal 1990; Bird \& Beers 1993) to
analyze the velocity distribution of our sample. We applied the method
of the weighted gap analysis as discussed by Ribeiro \etal (1998; see
also Capelato \etal 2008) in order to remove interlopers and to identify
the main kinematical structures. Figure \ref{fig:A376histov} (inset) shows
the radial velocity distribution of the whole redshift sample, where the presence
of a very dominant kinematical structure is confirmed by the gap analysis.
This kinematical structure, which we identify for A376, is displayed in
the main part of Figure \ref{fig:A376histov}. It is constituted of 89
galaxies with radial velocities ranging between $12500$ and $16300\kms$,
with mean $\overline{V}_{rec} = 14241^{+151}_{-172}\kms$, corresponding
to redshift $z_{A376} = 0.047503$. The  velocity dispersion corrected following
Danese \etal (1980) is $\sigma_{corr}=830^{+122}_{-90}\kms$\footnote{In this
paper means and dispersions are given as biweighted estimates, see Beers \etal
1990. Error bars are 90\% confidence intervals and are calculated by bootstrap
re-sampling of 10000~subsamples of the velocity data}.

\begin{figure*}[htb] \centering
\includegraphics[width=13cm]{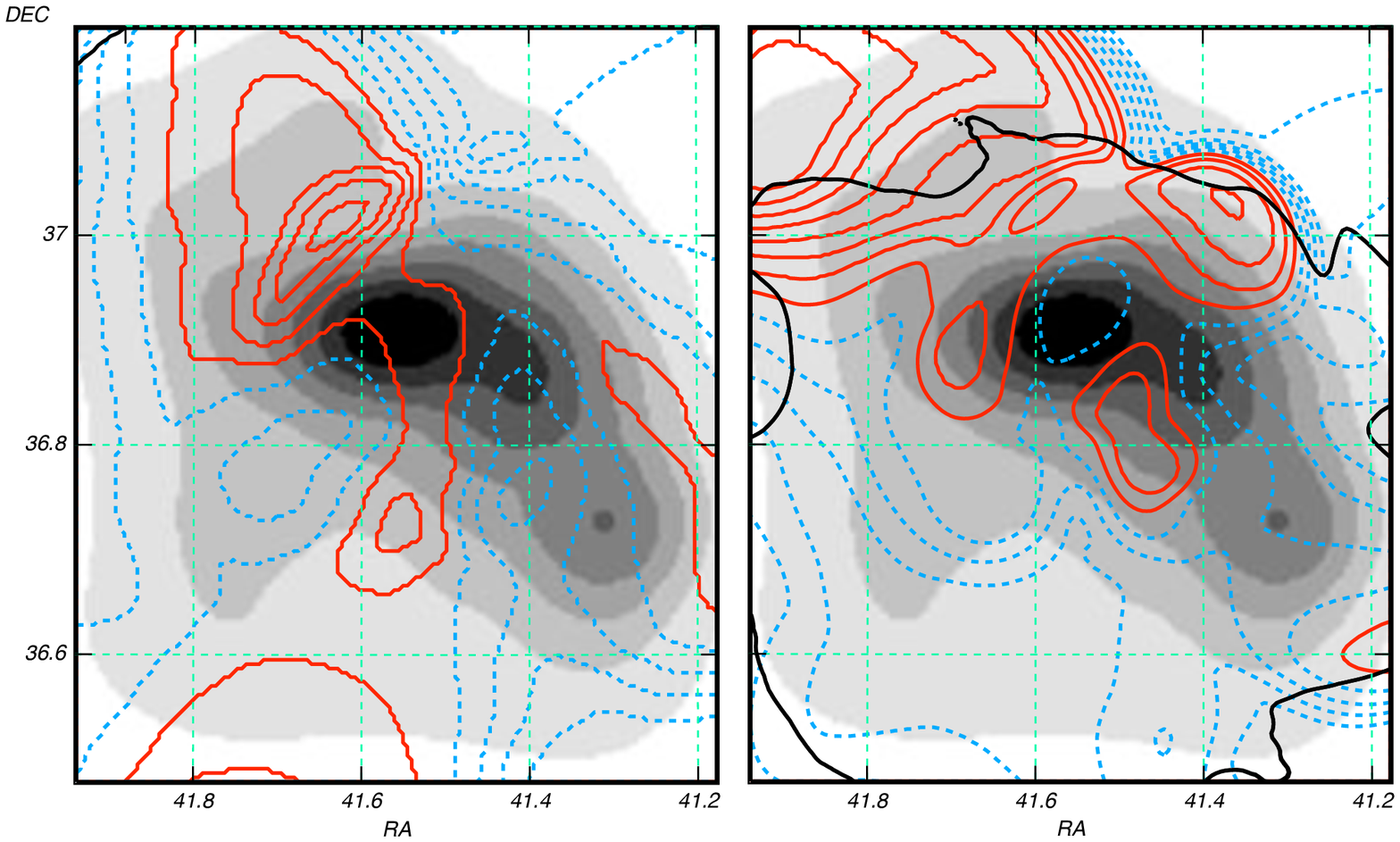}
\caption[]{Contours of equal local  mean velocity (left panel) and of local
mean dispersion velocity (right panel), superimposed on the gray-level AK
surface density image of the central region indicated in
Figure \ref{fig:A376map}. Higher values are shown with red continuous lines:
$\overline{V}_{rec}$ ($\Delta=200\kms$) between $14500$ and
$15500\kms$ (left) and $\overline{\sigma}$ ($\Delta=100\kms$) between
$900$ and $1600\kms$ (right); lower values with blue dotted lines:
$\overline{V}_{rec}$ between $12500$ and $14300\kms$ (left) and
$\overline{\sigma}$ between $400$ and $800\kms$ (right). The bold
black line in the left panel is the boundary of the region where
results are 3$\sigma-$significant after 10000 bootstraps of the data.
All samples are limited at 19.5 mag.}
\label{fig:pannels_mean_disp}
\end{figure*}

The ROSTAT routines detect a significant gap (indicated by an arrow in
Figure \ref{fig:A376histov}) $\Delta V \sim 100\kms$ in the  velocity
distribution sample at $V_{gap} \sim 14500\kms$ (3\% probability of being
drawn from an underlying normal distribution). To see if this reflects some
special feature of the galaxy distribution, in
Figure \ref{fig:pannels_mean_disp} (left  panel) we show the kernel
weighted  local mean velocity map for galaxies belonging to the kinematical
structure shown in Figure \ref{fig:A376histov}, together with their adaptative
kernel projected density map.

\begin{figure*}[htb]
\centering \includegraphics[width=8.6cm,angle=-90]{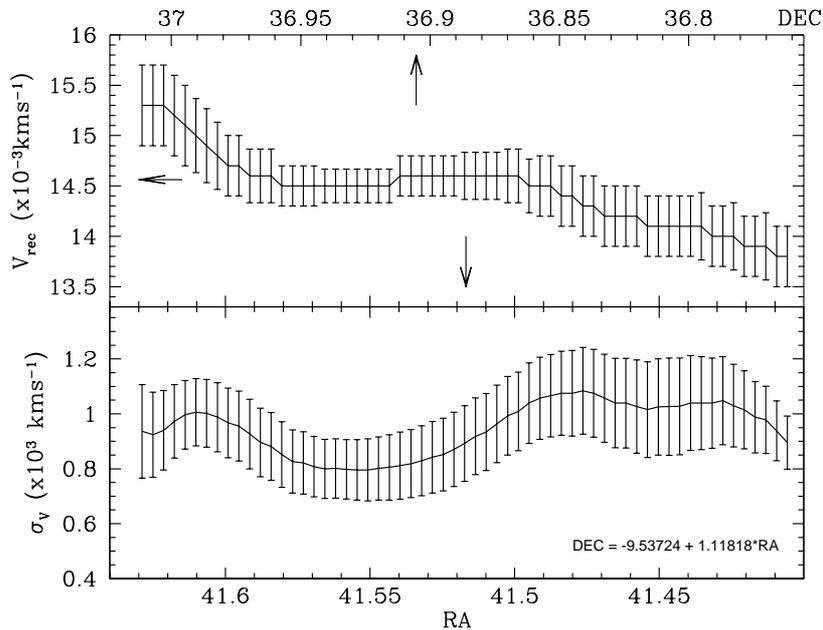}
\caption[]{The local mean velocity profile (upper panel) and the local
dispersion velocity  profile (lower  panel) taken  along a line showing the
highest velocity gradient (equation given in the lower panel). The arrows
indicate the position and velocity of the dominant E/D central cluster galaxy:
lower and upper arrows for the RA and DEC positions and the left upper arrow
for its velocity. Error bars are 1-$\sigma$ standard deviation of mean values
after 5000 data bootstraps.}
\label{fig:A376grad}
\end{figure*}

As seen from this figure, the high-velocity galaxies, $V >  V_{gap}$, are almost completely
concentrated to the north of the cluster center, characterizing an SW-NE
velocity gradient, possibly caused by a substructure being accreted by the
main cluster. This suggests that the velocity distribution is bimodal.
Indeed, the normality tests of ROSTAT already indicate that  the distribution
is assymetrically tailed, as is also apparent in Figure \ref{fig:A376histov}.
Figure \ref{fig:A376grad} shows the local mean velocity  profile taken along
the line of highest gradient. As seen, the velocity gradient only manifests
itself outside the core region of A376, which displays a uniform mean
velocity, very nearly the same as the E/D dominant galaxy (indicated by the
left arrow). This is interesting because, as noted in P03, when compared to the
cluster baricenter, the peculiar velocity of the dominant galaxy ($322 \kms$)
is only barely consistent with the distribution of peculiar velocities of cD
galaxies given by Oegerle \& Hill (2001). Our new analysis suggests that A376
is a far more complex structure in which only the main central core seems to
conform to the properties of a (classical) relaxed cluster, thought to be
centered on a large dominating spheroidal galaxy at rest relative to it.

Considering the velocity distribution of galaxies accordingly to their
morphological types, Figure \ref{fig:A376histov} shows the velocity
distribution of early type (E+S0: 61 objects) and late type (S+I: 28 objects)
galaxies, according to the classification given by Dressler (1980). Both
samples are limited at 19.5 mag. Their mean and dispersion velocities are
$\overline{V}_{early} = 14378 \pm 167\kms$ and $\sigma_{early} =
716^{+133}_{-96} \kms$, and for S+I types $\overline{V}_{late} =
13849^{+481}_{-390}\kms$, $\sigma_{late} = 966^{+316}_{-297}\kms$.
As observed in most clusters (Sodr\'e \etal 1989, Stein 1997, Carlberg \etal
1997, Adami \etal 1998), the  velocity  dispersion of  the late-type population
is larger than that of the early-type population. The results obtained in P03
remain unchanged with these new data.

\begin{figure*}[htb]
\centering \includegraphics[width=13cm]{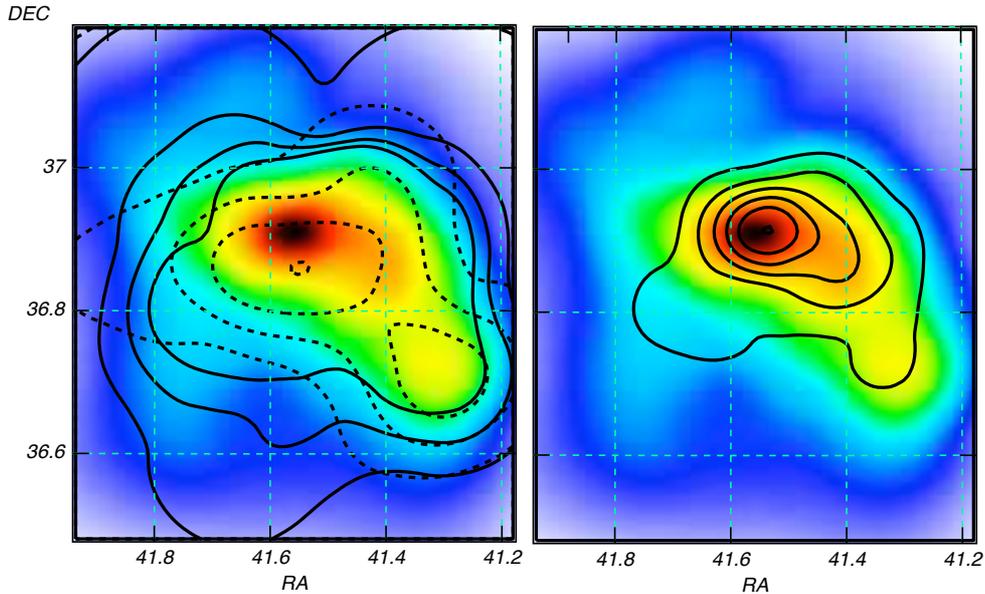}
\caption[]{Isopleths of the surface distribution of early (E+S0) galaxies
(continuous lines) and late (S+I) galaxies (dotted lines) superimposed on the AK
surface density image of the central region of Figure \ref{fig:A376map}. Left
panel: low-density isopleths $\Sigma_{5} < 1.4$, $\Delta_{5} = 0.35$. Right panel:
high-density isopleths $ 8.3 > \Sigma_{5} > 1.4$, $\Delta_{5} = 1.38$. $\Sigma_{5}$
is measured in units of $10^{-5} \mathrm{gal\cdot arcsec^{-2}}$.}
\label{fig:morpho_distrib}
\end{figure*}

To examine the relative contribution of early and late type galaxies to the
overall projected distribution of galaxies, we show the high and low density
isopleths of their distributions in Figure \ref{fig:morpho_distrib} (left and
right panels, respectively) superimposed on the AK surface density image of the
photometric sample, limited at 19.5 mag. Early type galaxies are largely dominant
over late-type galaxies by more than a factor 2 in number. As seen from this
figure, they are also much more concentrated (dense) in the center of the cluster,
by a factor $\sim$ 6. This is a clear demonstration of the effect of morphological
segregation acting locally in the cluster.

\subsection{Abell~970}

This cluster is extensively discussed in Sodr\'e \etal (2001, hereafter S01) and in
Lima Neto (2003), which have shown that this is a rather complex system. Including
the already published velocities (S01), 14 new redshifts for  a total of 83 have
been obtained in  the direction of Abell~970. 

\begin{figure}[htb]
\centering
\includegraphics[width=8.6cm]{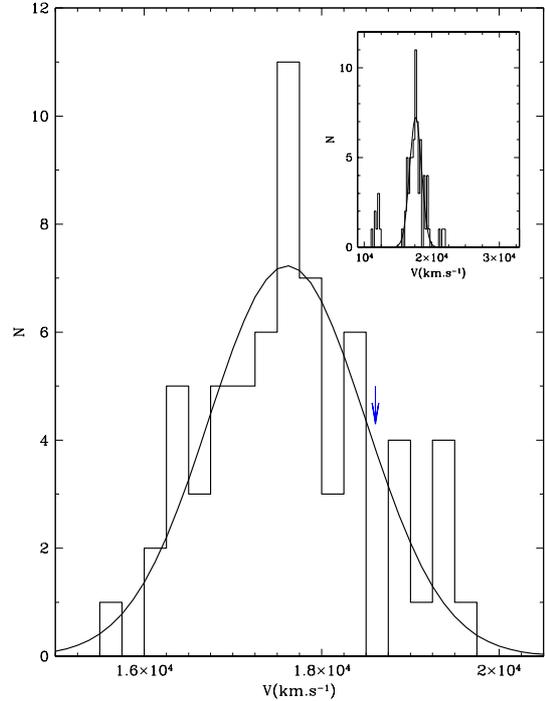}
\caption[]{The radial velocity distribution for the Abell~970 sample of
galaxies between $15000$ and $20000\kms$. The continuous curve shows the
Gaussian distribution as in Figure \ref{fig:A376histov}. The arrow at
$v \sim 18600\kms$ points to a significant gap found from the gap analysis.}
\label{fig:A970histov}
\end{figure}

As already done in S01, both the iterative gap analysis and the statistical tests
provided by ROSTAT were applied to remove contaminant interloppers in the redshift
sample. This has shown that the cluster radial velocities range between
$ \sim 15000\kms$ and $20000\kms$, with mean and dispersion velocities
$\overline{V}_{rec} = 17612^{+189}_{-182}\kms$ ($z_{clus} = 0.058747$) and
$\sigma =  881^{+144}_{-123}\kms$. The histogram of the velocity distribution is
displayed in Figure \ref{fig:A970histov}. 

These analyses have also shown a significant gap in the velocity distribution,
$V_{gap} \sim 18500\kms$, already reported by S01 (see Figure \ref{fig:A970histov}).
In that work it was suggested that the gap occured because of the bimodality of the
radial velocity distribution, signaling the state of non equilibrium of the cluster,
also proven by the presence of very compact clump of galaxies situated $\sim$ 8
arcmin ($544 \rm h_{70}^{-1}~Kpc$ at z = 0.059) NW of the BCG with mean velocity
$\overline{V}_{clump} = 19227\kms$. The new data presented here reinforces this
picture, since the new redshifts distribute everywhere outside the central $\sim$
1Mpc region of the cluster. The off-set of the X-ray emission distribution relative
to the galaxy distribution and the gas temperature and metal abundance gradients are
also strong evidence that A970 has suffered a recent merger with a subcluster or that
the NW substructure has recently passed through the center of A970 (see Lima Neto \etal,
2003 for details).

\subsection{Abell~1356}

Abell~1356 was classified as a Bautz-Morgan II-III morphology and richness class
R~=~1 (Abell \etal 1989), at redshift $z= 0.0698$ (Struble \& Rood 1999) based
only on 2~velocities. Up to now, the cluster has not been studied except by X-ray
observations. A1356 is within the sky area surveyed by the SDSS project which thus
provides data, both spectroscopic as photometric, and allowing detailed studies.
As seen in Table~1, 8 of the 10~redshift measurement we undertook in the field of
A1356 have also being targeted by SDSS. 

Jones \& Forman (1999) analyzed the \textit{Einstein} IPC X-ray image of  A1356.
The count rate is 0.0065~s$^{-1}$ in a region of $1h_{50}^{-1}~$Mpc radius
giving a luminosity $L_{\rm X} = 0.429 \times 10^{44}~$erg~s$^{-1}$ in the
[0.5--4.5]~keV band (notice  that they adopted the redshift of $z = 0.1167$,
based on Struble \& Rood 1987). They did not detect any cooling-flow and they were 
not able to derive any isocontour map of the X-ray emission.

An analysis of A1356 using ROSAT data was done by Romer \etal (2000) in the
context of the SHARC survey. They obtained a total flux
$f_{\rm  X} = 51.82 \times  10^{-14}~$erg~cm$^{-2}$~s$^{-1}$ in the
[0.5--2.0]~keV passband, which corresponds to a luminosity $L = 0.11
\times 10^{-44}~$erg~s$^{-1}$ assuming a temperature of 2~keV. ROSAT PSPC
observations have also been analyzed by Burke \etal (2003), also as part of the
SHARC survey. They find that A1356 has a flux $f_{\rm  X} =
57.51  \times 10^{-14}~$erg~cm$^{-2}$~s$^{-1}$  in  the [0.5--2.0]~keV
passband and luminosities $L_{X} = 0.12 \times 10^{44}~$erg~s$^{-1}$
in the [0.5--2.0]~keV and $L_{X} = 0.22 \times 10^{44}~$erg~s$^{-1}$
in the [0.3--3.5]~keV bands. Both these analyses of ROSAT data were done in an
automatic way as part of their pipeline for identifying distant cluster
candidates.

\begin{figure}[htb]
\centering
\includegraphics[width=8.6cm]{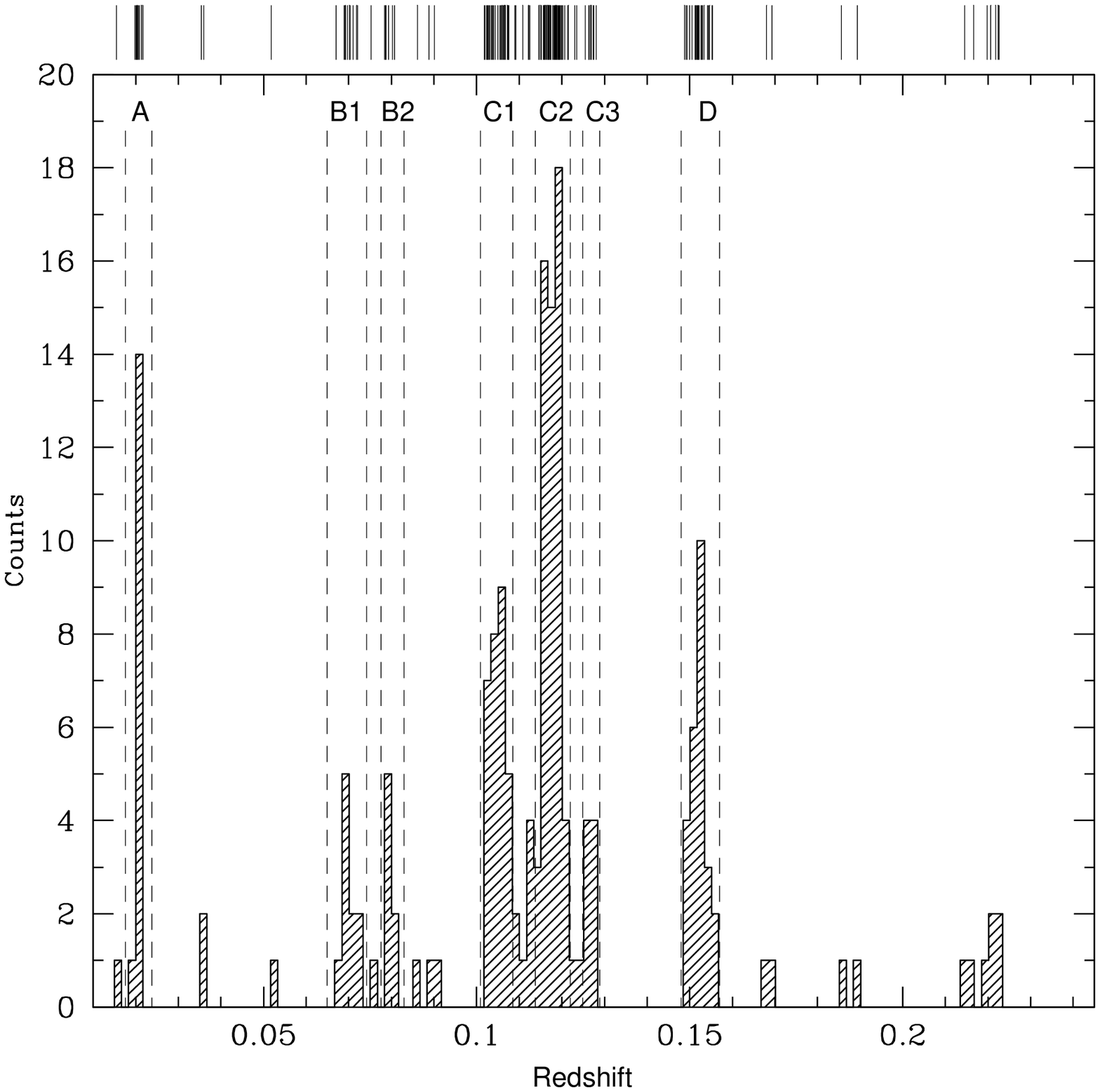}
\caption[]{The redshift distribution of a square region of $0.83^{\circ}$ side centered
on the nominal center of A1356 ($\rm 11^{h} 42^{m}28.8^{s} +10^{\circ}26^{'}21^{''}$); a
step corresponds to $500\kms$. The most significant kinematical structures identified
with gap analysis are indicated by different symbols within dashed vertical lines. The
bar code at the top of the plot shows the redshift coordinate of the entire sample.}
\label{fig:A1356_histoz}
\end{figure}

Figure \ref{fig:A1356_histoz} shows the redshift distribution of galaxies
within a square region of side $0.83^{\circ}$ centered on A1356, which
corresponds to $\rm 4 h_{70}^{-1} Mpc$ at z$ = 0.07$, the nominal redshift of
the cluster. Data is from the SDSS database added to our own measurements
(only 2~new redshifts, see Table~1) and limited to r = 22 mag. We iteratively
applied the method of weighted gap analysis (see Section 3.1) to identifify the
main kinematical structures in this distributions. These are denoted by letters
in Figure \ref{fig:A1356_histoz}, and their main properties are displayed in
Table 2. A1356, nominally at $z \simeq 0.07$, should  correspond to structure
$B1$, with a dispersion velocity $\sigma_{B1} = 384\kms$, which is more
characteristic of a sparse group than of a rich cluster of galaxies.

\begin{figure}[!htb]
\centering
\includegraphics[width=8.6cm]{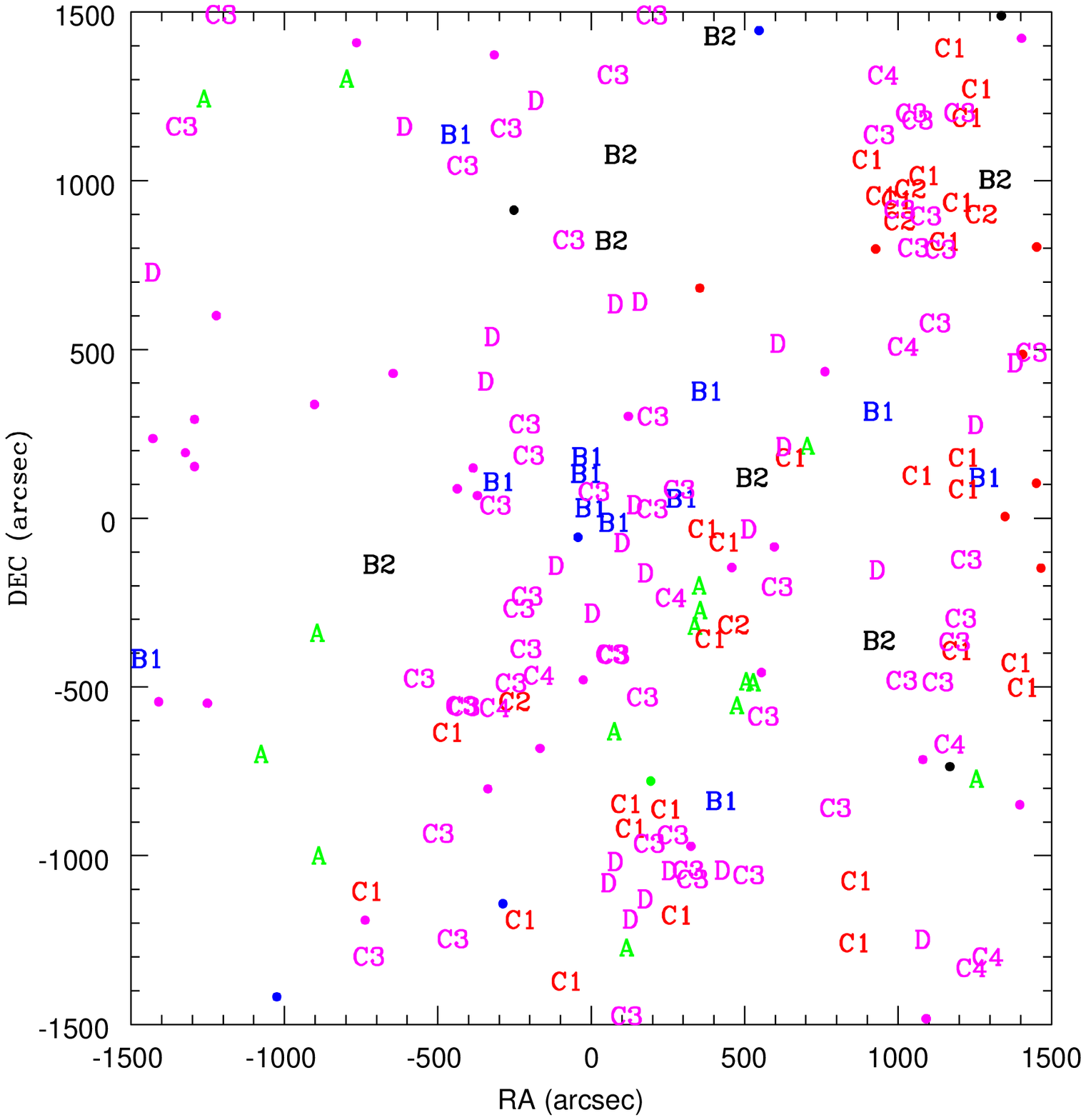}
\caption[]{The projected distribution of galaxies in the field of A1356. Different
symbols correspond to different kinematical structures displayed in Figure
\ref{fig:A1356_histoz} and in Table~2. Filled circles are galaxies outside of any
kinematical structure. The square region displayed in this figure is centered on the
nominal center of A1356 with a side equivalent to $\rm 4 h_{70}^{-1} Mpc$ at z= 0.07,
the nominal redshift of the cluster.}
\label{fig:A1356_XY}
\end{figure}

Figure \ref{fig:A1356_XY} shows the projected positions of galaxies in this same
region. As seen, galaxies belonging to $B1$ seem to constitute an elongated SE-W
structure, concentrated around the nominal center of A1356. This confirms the
reality of this system, although not as a rich cluster of galaxies as it was
initially believed, but probably just as a sparse group. The hypothesis that
the two kinematical groups could belong to one single structure, with the detected
gaps interpreted as due to the incompleteness of the sample, should be disregarded
for, given the distance modulus of 37.5, at $z = 0.07$, and assuming $M_{r}^{*} =
-20.94$, the mean value for rich clusters (Popesso et al, 2006), we get
$r^{*} = 16.6 $, well within the completeness limit of the SDSS spectrocopic survey
($r_{lim} \sim 17$).

\begin{figure*}[!htb]
\centering
\includegraphics[width=10.0cm]{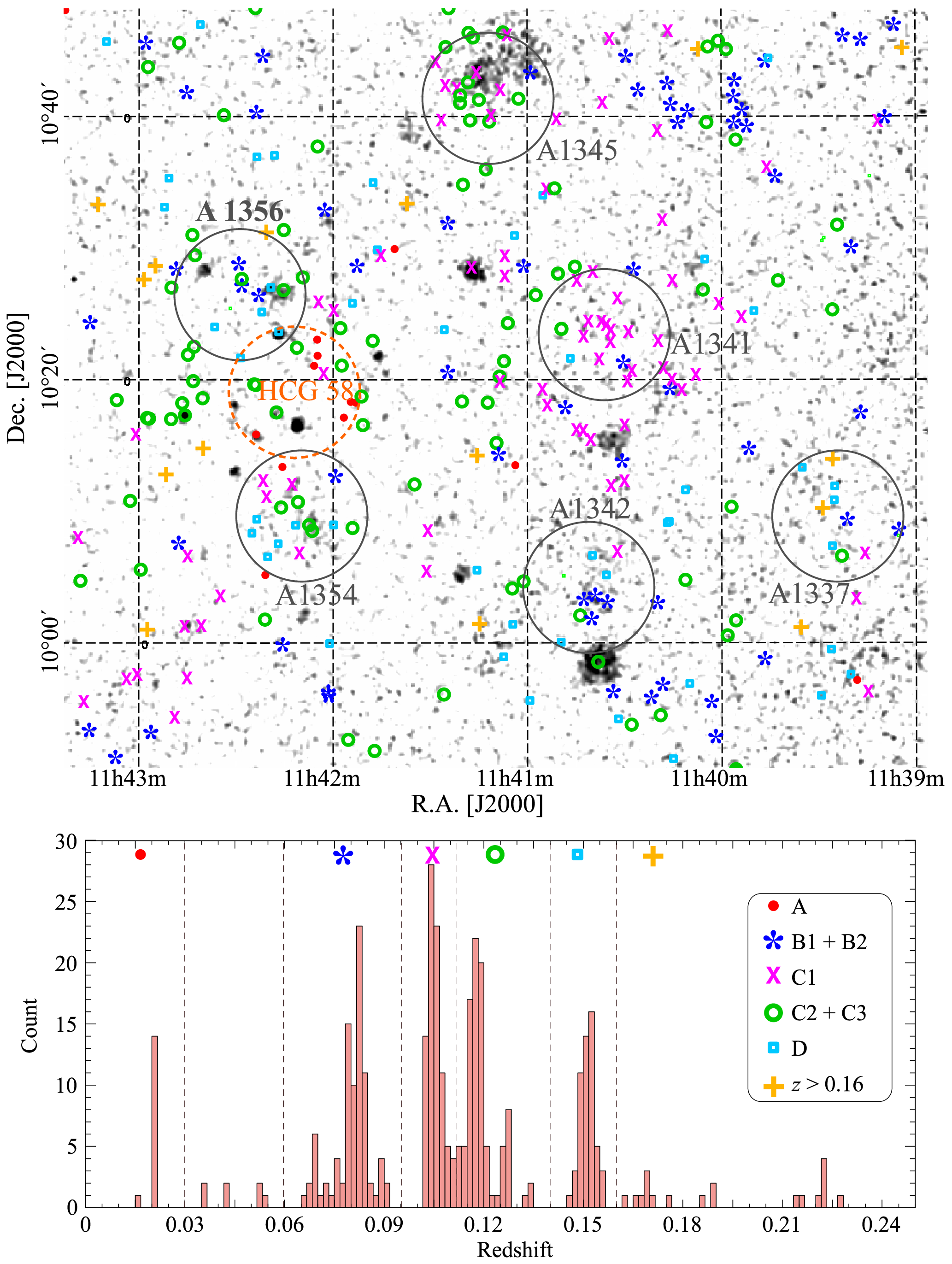}
\caption[]{ROSAT PSPC 0.1--2.4~keV image of the $\sim 30$~arcmin radius around A1356.
The large blue circles (4 arcmin radii) correspond to Abell clusters as indicated; the
large dashed circle, with 5~arcmin radius, corresponds to HCG~58. The smaller symbols
represent galaxies with measured velocities, each symbol corresponding to a redshift
range as indicated in the figure. There is no detected diffuse emission associated
with A1356.}
\label{fig:A1356_field}
\end{figure*}

With the exception of a  relatively important concentration at NW of the field
displayed in  Figure \ref{fig:A1356_XY}, which should be associated to the Abell cluster
A1345 (see  below), the other kinematical groups displayed in Table~2 do not appear to
be clearly concentrated on their projected surface distribution. It should be stressed
that, as for the results displayed in this Table, no single one of these groups have
dispersion velocity, which could characterize a rich cluster of galaxies, expected to
have, typically, $\sigma_{rich~cluster} > 800\kms$. In fact, as seen below (see Figure
\ref{fig:A1356_field}), no diffuse X-ray emission has been detected in the whole region.

\begin{table*}[htb]
\caption[]{Kinematical groups projected in the field of A1356}
\begin{tabular}{ccccc}
\hline
\hline
ID & Redshift range &  $\overline{z}$  &  $\sigma$ ($\rm km s^{-1}$) & no. of gal \\
\hline
A   &  $0.020 < z <  0.022$ & 0.02053 & $155^{-46}_{+59}$ & 15 \\
B1 &  $0.067 < z <  0.072$ & 0.06988 & $384^{-156}_{+186}$ & 12 \\
B2 &  $0.078 < z <  0.081$ & 0.07932 & $253^{-149}_{+199}$ & 7 \\
C1 &  $0.102 < z <  0.108$ & 0.10485 & $524^{-58}_{+86}$ & 29 \\
C2 &  $0.115 < z <  0.122$ & 0.11761 & $475^{-57}_{+69}$& 56 \\
C3 &  $0.126 < z <  0.128$& 0.12688  & $63^{-28}_{+22}$ & 8 \\
D   &  $0.149 < z <  0.155$& 0.15210  & $491^{-123}_{+117}$ & 25 \\
\hline
\end{tabular}
\end{table*}

\begin{figure}[!htb]
\centering
\includegraphics[width=8.6cm]{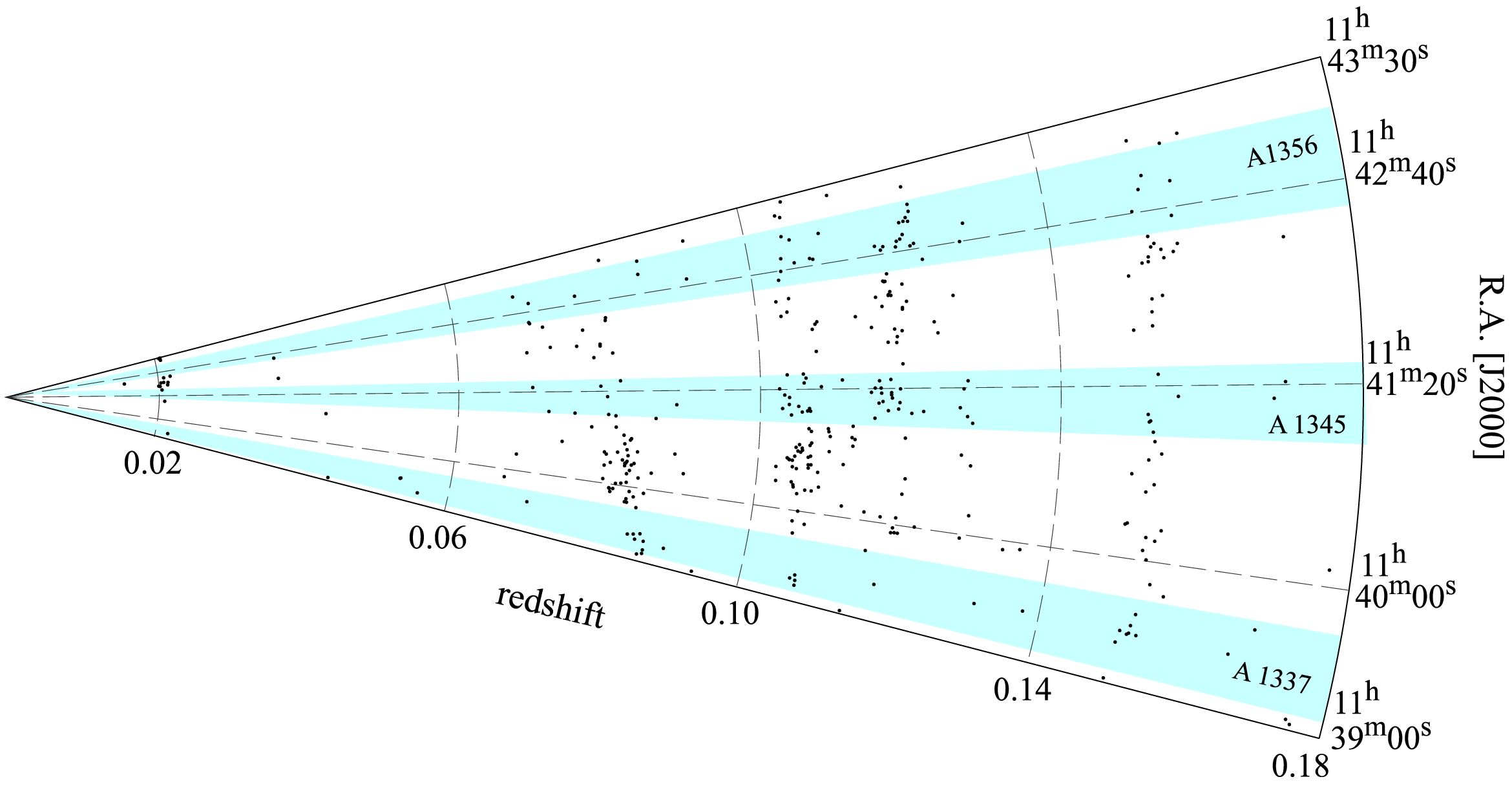}
\centering
\includegraphics[width=8.6cm]{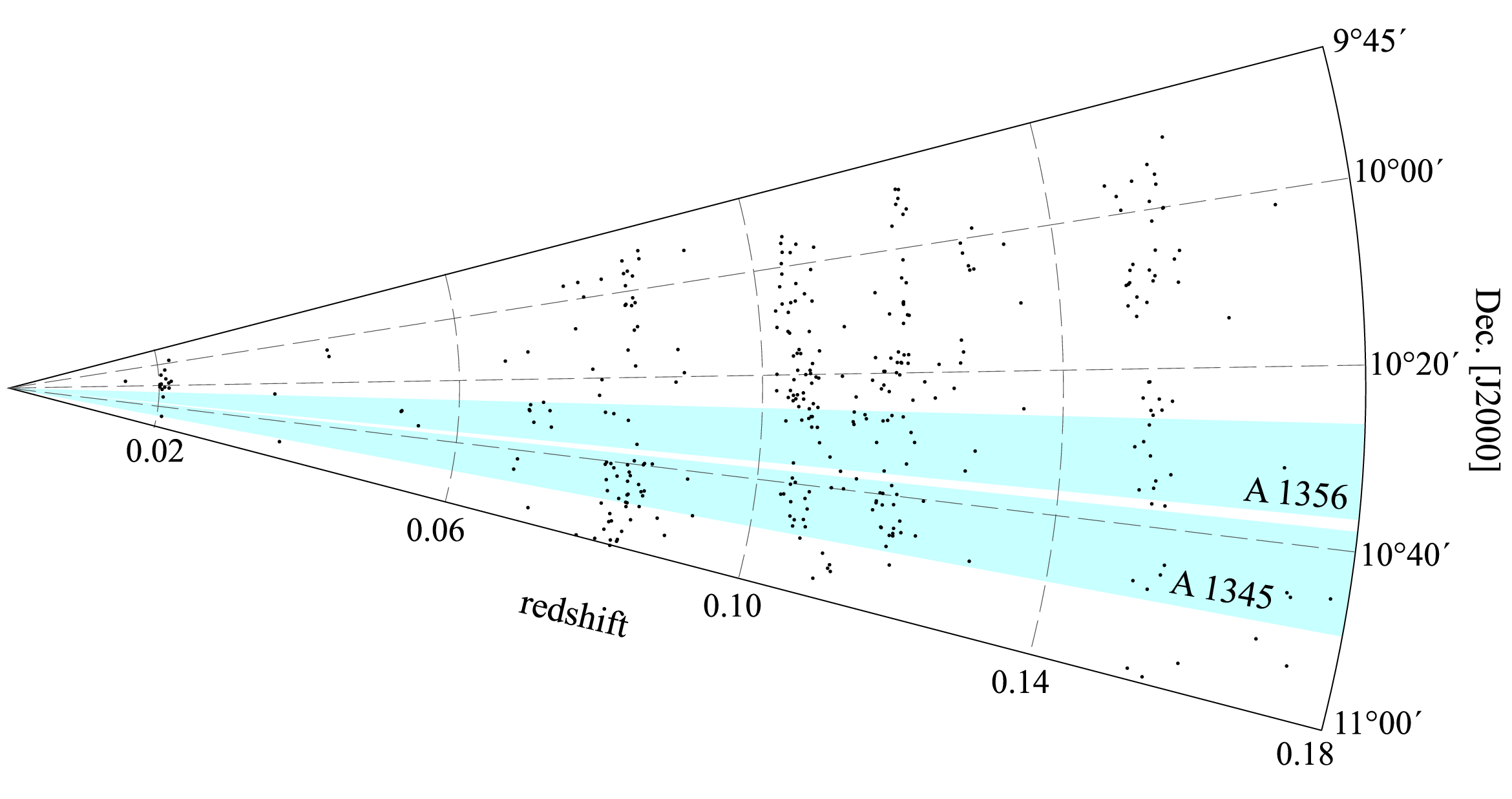}
\caption[]{Wedge diagrams in R.A. (left) and DEC (right) for the field
of cluster A1356. The shaded areas show the $11.5 \rm arcmin$ aperture
l.o.s view  cones for some of  the clusters present in  the field (see
also Figure \ref{fig:A1356_field})}
\label{fig:A1356_wedge}
\end{figure}

Figure \ref{fig:A1356_field} plots the positions of galaxies within the velocity peaks
from Table~2 over an X-ray image obtained with ROSAT PSPC in the 0.1--2.4~keV band. We
also plot  the positions of the clusters in the vicinity of A1356. A1345 is the most
evident one at 24~arcmin NW with velocities corresponding to the 3rd and 4th
velocity-peak objects of Table~2. This cluster is composed of 2~main structures on the
line of sight as shown on the wedge diagrams. The other noticeable clusterings are those
of galaxies with  $0.087 < z < 0.112$ at the position of A1341 and foreground objects,
$0.02 < z < 0.065$ that are part of the HCG~58 (Hickson compact group). Figure
\ref{fig:A1356_wedge} shows the redshift wedge diagrams for this sample, in both R.A.
and Dec projections. The projections of an l.o.s view cone with $\sim 11\rm arcmin$
aperture for  A1356 ($\rm \sim 0.9h_{70}^{-1} Mpc$ @ z= 0.07) are also shown in these
figures.

In their study of the distribution of Abell clusters and superclusters, Einasto \etal
(1996,2001) defined  the  \textit{very rich} ``Leo-Virgo'' supercluster at a mean redshift
$\rm  z_{SCl} = 0.112$ as composed of 8~Abell clusters, 6 of them with ``known distances'':
A1341 (z= 0.1049), A1342 (z= 0.1061), A1345 (z= 0.1095), A1354 (z= 0.1178), A1372
(z=0.1126), and  A1356 (z= 0.117); and 2~other clusters with only ``estimated'' distances,
A1379 and A1435. However, our analysis above pointed out that A1356 should be considered
as just a foreground \textit{group} of galaxies, located at z = 0.0689, thus giving no
contribution to the Leo-Virgo supercluster. In fact, the field around A1356 seems highly
contaminated by background galaxies, with a substantial fraction of them belonging at
redshifts of the Leo-Virgo supercluster (groups $C1$, $C2$, and probably $C3$ of Table~2),
and this should be the reason for the erroneous description of the cluster. Note that the
cluster A1435 has a redshift z = 0.062 (NED) and should also be considered as a foreground
cluster.

\begin{figure*}[!htb]
\centering
\includegraphics[width=12.0cm]{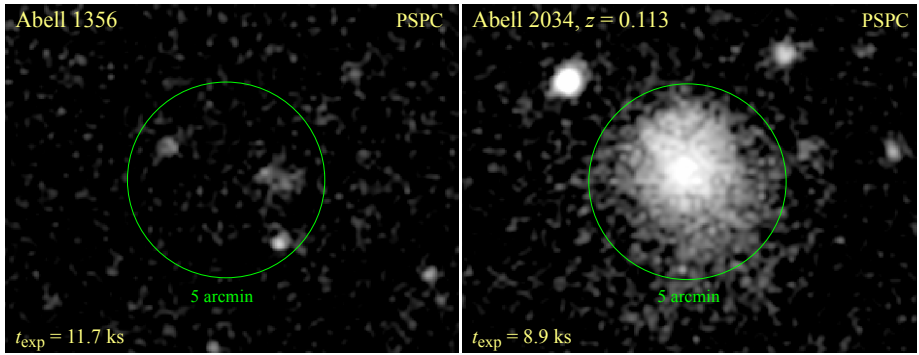}
\caption[]{Comparison between the images  of A1356 and A2034. A2034 is a Bautz-Morgan II-III
cluster of richness class 2 at $z=0.113$ observed for 8.9~ks with the ROSAT PSPC. If A1356 were a real cluster, its image should look very similar to the image of A2043.}
\label{fig:A1356xA2034}
\end{figure*}

The ROSAT PSPC X-ray image is rather shallow, with an exposure time of 11.7~ks, but it
should detect an Abell cluster. To verify that indeed the PSPC could detect extended
diffuse emission from a cluster up to $z \approx 0.1$, we selected another Abell cluster
that was observed by the PSPC with similar conditions, Abell~2034. It is a $z = 0.113$
cluster classified as a II-III B-M class and richness class~2. Although it is somewhat
richer than A1356 (which is classified as richness class~1), it was observed with a
shorter exposure time, 8.9~ks compared to 11.7~ks for A1356. We constructed an image for
each cluster in exactly the same way and show them in Figure \ref{fig:A1356xA2034}. The
difference is striking, so we can conclude that there is no sign of diffuse emission from
A1356. If the redshift of A1356 is indeed $\sim 0.07$, as we pointed out above, rather than
$\sim 0.11$, the discrepancy between the  images would be even stronger.

\subsection{Abell~2244}

Abell~2244 has a I-II morphology in the Bautz-Morgan classification and a richness class
R~=~2 (Abell \etal 1989). It is a cD or D-galaxy dominated cluster, and its brightest
cluster galaxy has an absolute magnitude of~-22.0 and a diameter of 98~kpc at the
25~Vmag arcsec$^{-2}$ isophote (Schombert \etal 1989). These authors give a redshift
$\overline{z}= 0.0968 \pm 0.0008$ with a dispersion $\sigma= 0.00414$. They note the
presence of a very near companion to the cD galaxy at a distance of 3~arcsec (4~kpc)
and a velocity difference 50$\kms$, which could be a final stage of merging. They also
suggest a subclustering in the line of sight from a set of 18~pairs of galaxies where 
only one is bound. A second structure has been evidenced by Miller \etal (2005) at a
redshift $z= 0.1024$.

\begin{figure}[!htb]
\centering
\includegraphics[width=8.6cm]{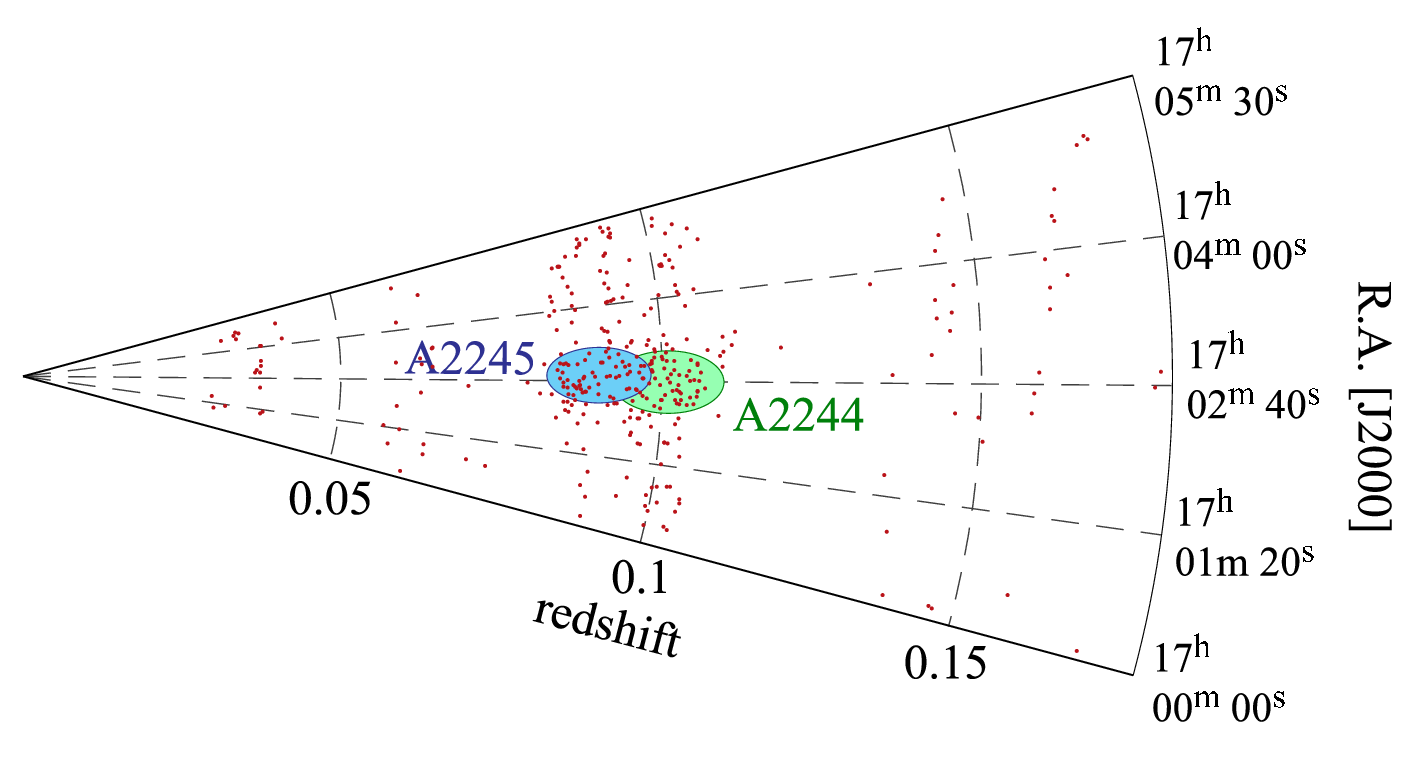}
\centering
\includegraphics[width=8.6cm]{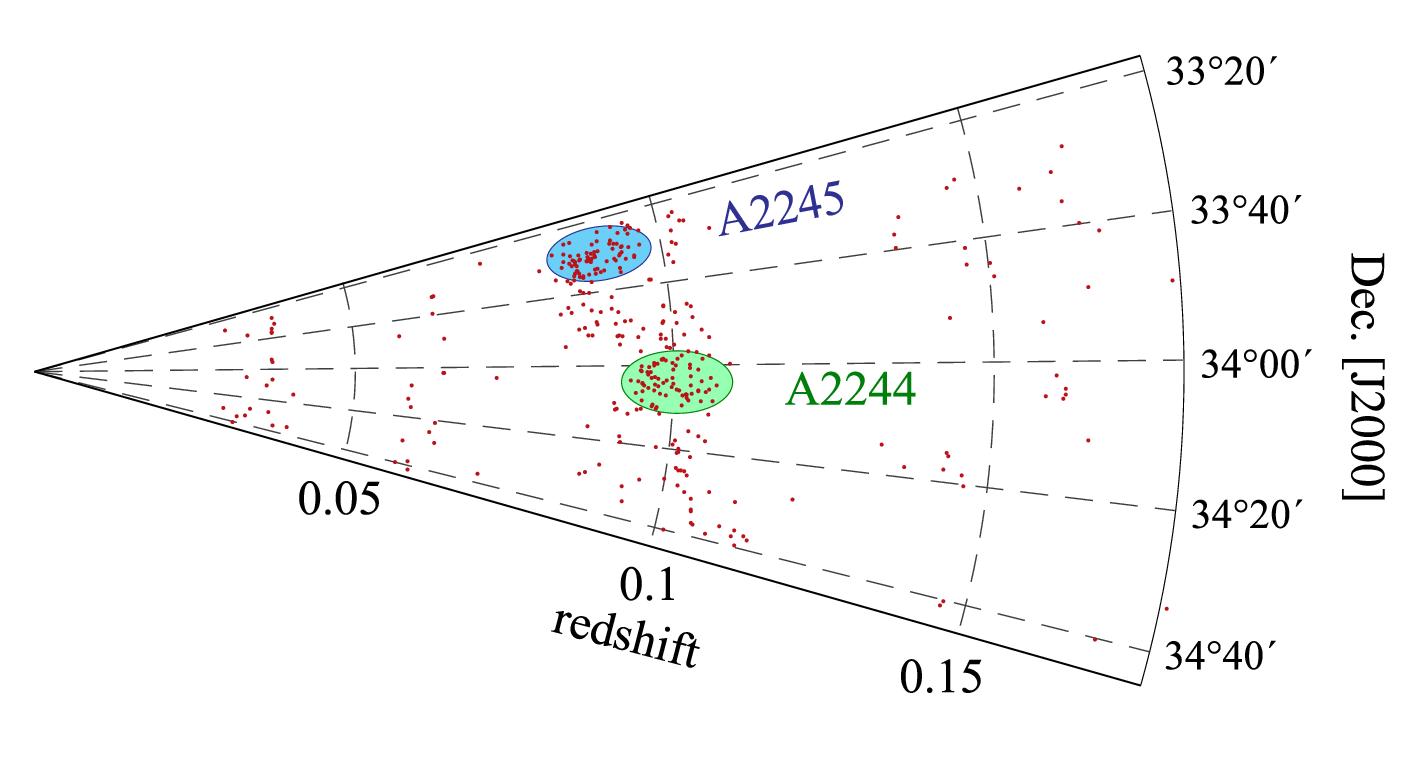}
\caption[]{Wedge diagrams in R.A. and DEC of the cluster A2244.
The ellipses are 3~Mpc in radius, perpendicular to the line of sight
and 1000$\kms$ in radius along the l.o.s.}
\label{fig:A2244_wedge}
\end{figure}

\begin{figure*}[!htb]
\centering
\includegraphics[width=8.6cm, angle=-90]{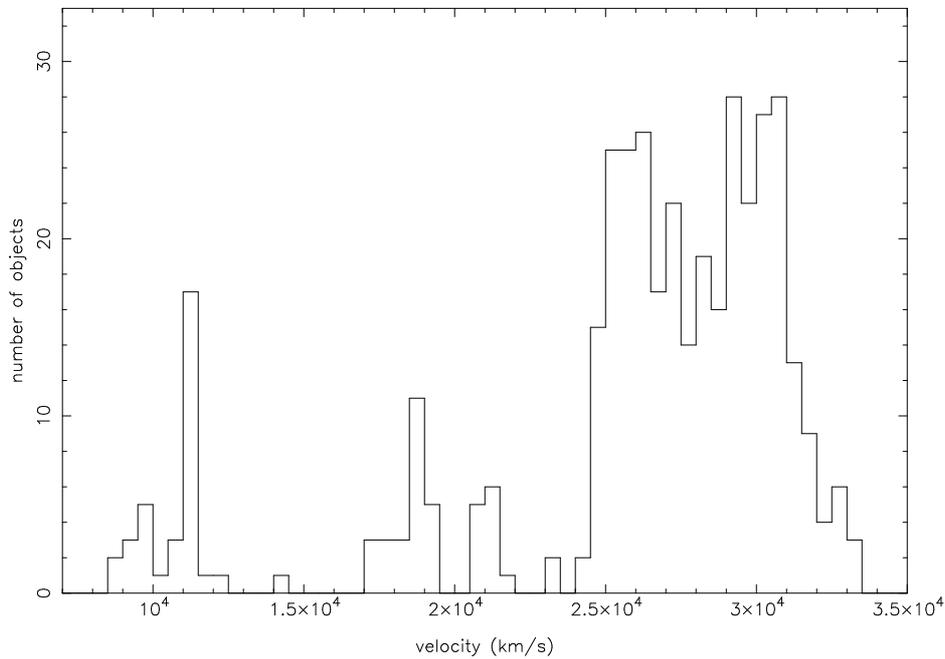}
\caption[]{The radial velocity distribution between $8000$ and $35000\kms$
(step $500\kms$) of the Abell~2244 and Abell~2245 sample of galaxies.}
\label{fig:A2244_histov}
\end{figure*}

The wedge diagrams in R.A. and Dec. of the 417~velocities collected in a 40~arcmin
radius (4.64~Mpc) from the NED database completed with our results are displayed in
Figure \ref{fig:A2244_wedge}. The positions of the two clusters A2244 and A2245 are
represented with ellipses of 3~Mpc in radius perpendicular to the line of sight and
$1000~\kms$ in radius along  the l.o.s. Instead of a  double cluster as
quoted by Struble \& Rood (1999), the two clusters belong to a much larger structure
visible in R.A. and Dec. at an average velocity of $28500~\kms$, which is associated
to a supercluster by Einasto \etal (2001), including A2249. The velocity histogram of
Figure \ref{fig:A2244_histov} shows 2~main peaks corresponding to the main velocities
of A2244 and A2245. From the galaxies contained  in each ellipse of Figure
\ref{fig:A2244_wedge} we obtain $C_{BI}= 29867^{+79}_{-86}\kms$ and $S_{BI} =
965^{+63}_{-66}\kms$ (110~galaxies) for  A2244 and $C_{BI} = 26197^{+102}_{-108}\kms$
and $S_{BI} = 992^{+57}_{-75}\kms$ (94~galaxies) for A2245. Rines and Diaferio (2006)
obtained for the same clusters respectively $\overline{v}= 29890\kms$, $\sigma =
981^{+95}_{-74}\kms$ and $\overline{v} = 26022 \kms$, $\sigma = 952^{+89}_{-70}\kms$
respectively. For both clusters, there is a very good agreement with Rines and
Diaferio (2006), who calculate the velocity dispersion profile within about $r_{200}$
following Danese \etal (1980). Note also the presence of two foregorund structures with
prominent peaks at $11191~\kms$ and $18461~\kms$.

\begin{figure}[!htb]
\centering
\includegraphics[width=8.6cm]{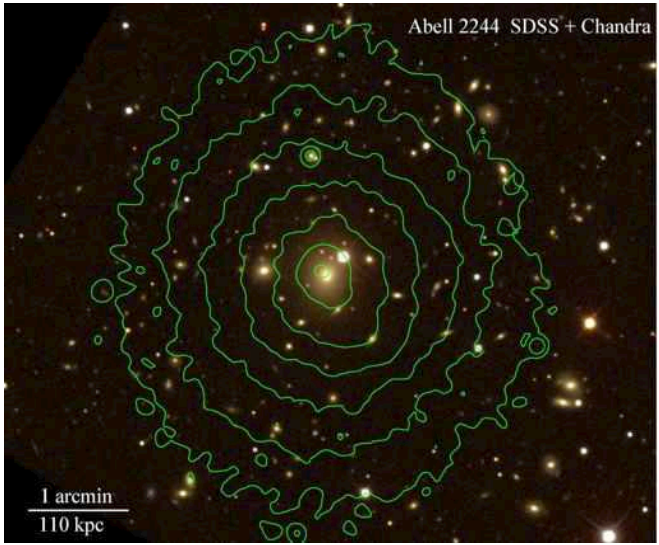}
\caption[]{{\it Chandra} X-ray contours superimposed to the gri Sloan
image in the central part of A2244.}
\label{fig:A2244_X}
\end{figure}

Abell~2244  was observed  in X-rays  by {\it  Chandra}  (Donahue \etal 2005). It is
nearly isothermal with $kT= 5.5 \pm 0.5$~keV at every radius $\prec 4$~arcmin.
Figure \ref{fig:A2244_X} shows the gri Sloan image and the {\it Chandra} contours in
the central part of A2244. A small offset between the center of the cD galaxy and the
X-ray peak is visible as a sharp edge on the X-ray image (a cold front or a shock front)
about 10~arcsec NE from the center, possibly indicating a movement in that direction.
Quoting Donahue \etal (2005), there is no evidence in the X-ray surface brightness
map for fossil X-ray cavities produced by a relatively recent episode of AGN heating. No
interaction seems to be present between A2244 and A2245.

\section{Summary}

In this paper we presented a set of 80~new radial velocities in the direction of
4~Abell clusters of galaxies: Abell~376, Abell~970, Abell~1356, and Abell~2244. 

For A376 we obtained an improved mean velocity value $\overline{V}_{rec} = 14241\kms$
and velocity dispersion $\sigma = 830\kms$. The new data suggest that A376 displays a
complex structure with evidence of bimodality in the radial velocity distribution where
only the main central core seems to conform to the expected features of a relaxed cluster.
The effect of morphological segregation acting locally in  the cluster is clearly seen both
in the surface distribution of galaxies and in their radial velocity distribution.

For A970, we have $\overline{V}_{rec} = 17612~\kms$ and $\sigma = 881~\kms$. Previous
analyses have shown that the cluster has substructures and is out of dynamical equilibrium.
The new data presented here confirms this conclusion.
 
We analyze the cluster  A1356 for the first time. We derive a new velocity value
$\overline{V}_{rec} = 20986~\kms$ with $\sigma = 384~\kms$. This cluster would not be a
member of  the ``Leo-Virgo'' supercluster as well as the cluster A1435 at
$\overline{V}_{rec} = 18588~\kms$.  

We obtain for A2244 $\overline{V}_{rec} = 29867\kms$ and $\sigma =  965\kms$ and for
A2245 $\overline{V}_{rec}  = 26197\kms$ and $\sigma = 992\kms$. These two clusters are
members of a possible supercluster including A2249. From optical and X-ray data, these
A2244 and A2245 show no sign of interaction.

\acknowledgements{We thank the Haute-Provence observatory staff for their
assistance during the observations. HVC, LSJ, and GBLN acknowledge the financial
support provided by FAPESP and CNPq. DP acknowledges IAG/USP for its
hospitality.}


\end{document}